\documentclass[fleqn,12pt]{wlscirep}
\usepackage{graphicx}
\usepackage{amsmath,amssymb}
\usepackage{color}
\usepackage{url}
\usepackage{xcolor}
\usepackage{eucal}
\usepackage{setspace}

\newcounter{firstbib}

\def\citeR{\cite}
\def\bibliographystyleR{\bibliographystyle}

\title{Ising model for melt ponds on Arctic sea ice}

\author[1]{Yi-Ping Ma}
\author[2]{Ivan Sudakov}
\author[3]{Courtenay Strong}
\author[4,*]{Kenneth M. Golden}

\affil[1]{Northumbria University, Department of Mathematics, Physics, and Electrical Engineering, Newcastle upon Tyne, NE1 8ST, UK}
\affil[2]{University of Dayton, Department of Physics, 300 College Park, SC 111, Dayton, Ohio 45469-2314, USA}
\affil[3]{University of Utah, Department of Atmospheric Sciences,
135 S 1460 E Rm 819, Salt Lake City, Utah, 84112-0102, USA}
\affil[4]{University of Utah, Department of Mathematics, 155 S 1400 E Rm 233,
         Salt Lake City, Utah 84112-0090, USA}
\affil[*]{email: golden@math.utah.edu}

\begin{document}
\maketitle

\noindent\sffamily\textbf{Perhaps the most iconic feature of melting Arctic sea ice is the formation of distinctive, complex ponds on its surface during late spring. The evolution of melt ponds and their geometrical characteristics determines the albedo of sea ice, a key parameter in climate modeling \cite{ScFe10,PoPeCo12,Polashenski:JGR:2012,Flocco:2012:JGR,SFFT14}. However, a theoretical understanding of this evolution, and predictions of geometrical features, have remained elusive. To address this fundamental problem in polar climate science, here we introduce
a two dimensional random field Ising model
for melt ponds. The ponds are identified as
metastable states
\cite{AJ85_PRB,GG87_PRB,PRT08_PRB} of the system,
where the binary spin variable
corresponds to the presence of melt water or ice on the
sea ice surface.
With only a minimal set of physical parameters, the model predictions agree very closely with observed power law scaling of the pond size distribution \cite{Perovich:JGR:2002} and critical length scale where melt ponds undergo a transition in fractal geometry \cite{HASPG12}.}

\begin{figure}
\centering
\begin{tabular}{cc}
\includegraphics[width=0.5\textwidth]{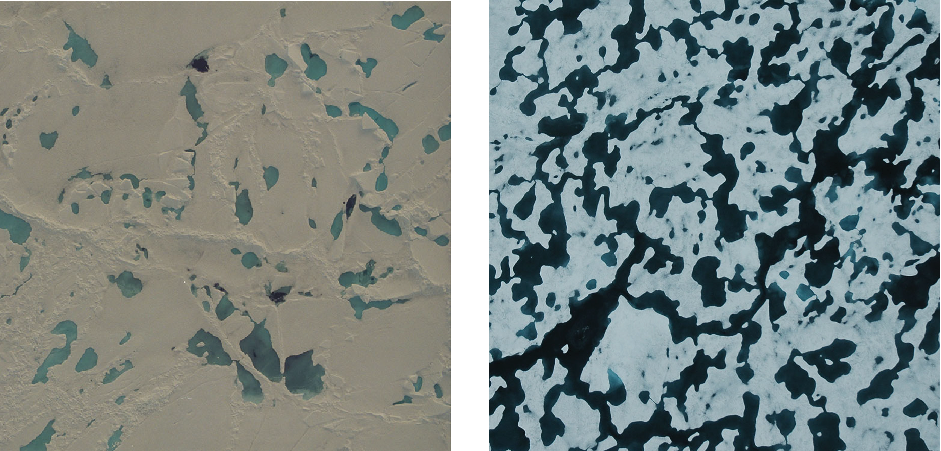} &
\includegraphics[width=0.25\textwidth]{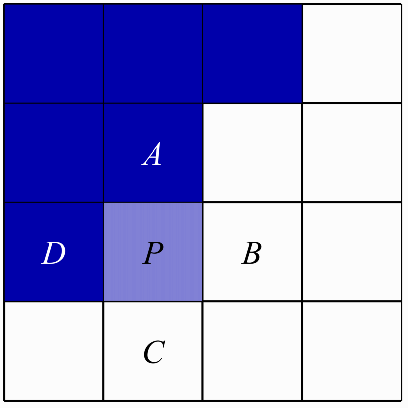} \\
(a) & (b)
\end{tabular}
\caption{ \textbf{Melt pond configurations and the
update step in Glauber dynamics.}
(a) Helicopter photos of melt ponds on Arctic sea ice in
the western Beaufort Sea (courtesy of D. Perovich).
On the left, each side of this 15 July 1998 photo is 826 meters;
on the right, each side of this 14 August 2005 photo is 193 meters.
(b) Illustration of an update step in Glauber single
spin-flip dynamics of the Ising model. Here each site $i$ is
assigned a pre-melt ice height $h_i$, and colored dark blue
for water ($s_i=+1$) and white for ice ($s_i=-1$). Site $P$,
to be updated, is adjacent to two water sites $A$ and $D$,
and two ice sites $B$ and $C$. To describe the tendency for water to fill troughs,
we require that $s_P=+1$ if $h_P<0$, and $-1$ otherwise.}
\label{fig:photo}
\end{figure}

\rmfamily
While snow and ice
reflect most incident sunlight,
melt ponds absorb most of it.
The ponds largely control
solar reflectance and transmittance of sea ice
\cite{ScFe10,PoPeCo12,Polashenski:JGR:2012,SFFT14}, which
in turn impact the heat and mass balances of
the ice cover and the partitioning of energy in the upper ocean
and lower atmosphere.
Typical pond configurations are shown in Fig.~\ref{fig:photo}(a).
It has been found \cite{Flocco:2012:JGR} that if a melt pond parameterization is included
in climate model simulations, then predicted September ice volume from 1990 to 2007
is nearly 40\% lower than in simulations which do \textit{not} incorporate ponds,
and is in much closer alignment with observations. Moreover, the yearly
Arctic sea ice minimum can be accurately forecasted from melt pond
area in spring \cite{SFFT14}.
The impact of melt pond evolution
extends into the biosphere as well \cite{Arrigo_S_2012,Nicolaus_GRL_2012},
where the ponds act as {\it windows} for light to shine into the upper ocean,
affecting Arctic marine ecology.

There has been significant progress on the development of
numerical models of melt pond evolution
\cite{ScFe10,Polashenski:JGR:2012,Flocco:2012:JGR,SFFT14}.
However, a fundamental theory of melting sea ice
which accounts for observed pond characteristics has been lacking.
Here we look toward statistical physics,
and the Ising model in particular \cite{YEO92,Christensen:2005},
to develop such a theory.
We envision surface
patches or {\it pixels} of ice or melt water as
collectively influenced by an external forcing field, and
interacting only with their nearest neighbors.

A central issue in climate science is {\it linkage of scales} $-$
that is, how can knowledge of small scale local interactions
be used to predict {\it macroscopic}
behavior relevant
to large scale, coarse-grained models?
This is the type of question that is addressed in
statistical physics \cite{Christensen:2005,YEO92} and
homogenization for composite materials
\cite{Milton:2002:TOC,Torquato:RHM-02}, where powerful methods of calculating
macroscopic behavior from ``microscopic'' laws
or microstructural information have been developed.
Indeed, an Ising model for tropical convection
was developed \cite{MaKh02} to represent unresolved features
in the atmosphere. Here we employ such methods
to represent a critically important unresolved feature in the polar marine environment.

First, we recall the most general form of the Ising free energy,
\begin{equation}\label{eq:Ising-general}
\mathcal{H}=-\sum_iH_is_i-\sum_{\langle i,j\rangle}J_{ij}s_is_j,
\end{equation}
where $i$ ranges over a two dimensional square lattice
with periodic boundary conditions, and $\langle i,j\rangle$
denotes nearest neighbors. In our model the state variable is a
binary (or spin) variable $s_i$ such that $s_i=+1$ corresponds
to absorptive melt water on the surface of our pixelated
model sea ice floe and $s_i=-1$ corresponds instead
to reflective ice or snow on the surface.
The parameters $H_i$ and $J_{ij}$
represent the external magnetic field and coupling constants,
respectively.
In addition, a temperature $T$ can be defined which controls
the strength of thermal fluctuations, but here we set $T=0$ assuming
that environmental noise does not significantly influence melt pond formation.

To describe nontrivial spin clustering at zero temperature,
the $H_i$ and/or $J_{ij}$
are chosen as random variables; the resulting models are collectively
known as disordered Ising models \cite{Natter}.
In particular, one recovers the classical random
field Ising model (RFIM) if the $H_i$ are independent
random variables and the $J_{ij}=J$ are constant.
At zero temperature, the system is usually assumed to follow Glauber single spin-flip dynamics \cite{KrReBe10_Book}: at each update step, the flip is accepted if $\mathcal{H}$ decreases and rejected if $\mathcal{H}$ increases. The system eventually converges to a local minimum of $\mathcal{H}$, known as a metastable state.

Metastable states are especially relevant to physical systems near phase transitions, including supercooled liquids \cite{BH00_PRL} and atmospheric aerosol particles \cite{RSLC89_Nature}.  For disordered Ising models they have been realized experimentally in, for example, doped manganites \cite{MoMaFeYuDa00} and colossal magnetoresistive manganites \cite{WIHPCD06_NatMat}. Despite their importance, metastable states are not completely understood theoretically \cite{KrReBe10_Book}, with analytical results largely restricted to 1D \cite{DeGa86_JPhys} and many intricate issues remaining in 2D \cite{NeSt99_PRE}.

The key factor controlling melt pond configurations is the pre-melt ice topography, represented by random variables $h_i$. In the spirit of creating order from disorder, these variables are assumed to be independent Gaussian with zero mean and unit variance. The lattice constant $a=0.85$ m is specified as the length scale above which important spatially correlated fluctuations occur in the power spectrum of sea ice topography (see Supplementary Methods). We use the following update rule for Glauber dynamics, depending on whether there is a majority among the four neighbors of a chosen site. If a majority exists, the site is updated
to align with the majority because of heat diffusion between
neighboring sites. Otherwise, we introduce a
\textit{tiebreaker} rule that describes the tendency for water to fill troughs:
the chosen site is updated to ice if its pre-melt ice height
is positive, and water otherwise; see Fig.~\ref{fig:photo}(b).
Note that this update rule does not depend
on any parameters other than $h_i$.

The above update rule can be restated as minimizing the classical RFIM free energy \cite{AJ85_PRB,GG87_PRB,PRT08_PRB}
\begin{equation}\label{eq:H-RFIM}
\mathcal{H}=\sum_i(h_i-H)s_i-\sum_{\langle i,j\rangle}Js_is_j,
\end{equation}
with the uniformly applied field $H=0$ and the coupling constant $J\rightarrow+\infty$; see Supplementary Methods for a brief discussion of the $H\neq0$ case. To facilitate comparison with geophysical observations, the order parameter will be taken as the pond fraction $F$, which is defined as the fraction of up-spins and therefore related to the magnetization $M$ by $F=(M+1)/2$. At $J=0$, there is a unique metastable state given by $s_i=+1$ if $h_i<H$, and $s_i=-1$ if $h_i>H$.
This
process can only yield
the correct melt pond geometry if the random variables
$h_i$ are highly correlated \cite{Bowen_JFG_2016}.
As $J$ increases, metastable states appear \cite{WuMa05_PRL} at a wider range of
pond fractions, with the entire range $F\in[0,1]$
covered for large enough $J$.

\begin{figure}
\centering
\begin{tabular}{ccc}
\includegraphics[width=0.24\textwidth]{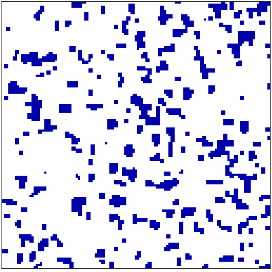} &
\includegraphics[width=0.24\textwidth]{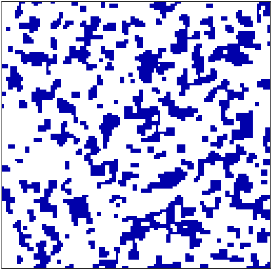} &
\includegraphics[width=0.24\textwidth]{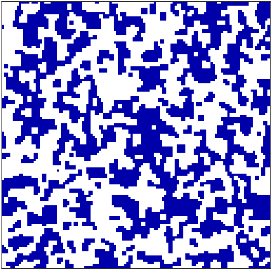} \\
(a) & (b) & (c)
\end{tabular}
\caption{\textbf{Melt ponds as metastable \textit{islands} of like spins in our random
field Ising model.}
Simulation results are shown for metastable states
of the RFIM at $H=0$ and $J=5$. The output spin configurations are shown
on a $128\times128$ portion of the $1024\times1024$ lattice
with (a) $F_{out}=0.15$; (b) $F_{out}=0.30$; (c) $F_{out}=0.45$.
Pixels are colored blue for water ($s_i=+1$) and white for ice ($s_i=-1$).}
\label{fig:MEPO_det_evo}
\end{figure}

Below we present numerical results for the zero temperature Glauber dynamics of the RFIM, with $10$ Monte Carlo steps used for each simulation and the lattice size taken to be $1024\times1024$. The input spin configurations $s_i$ are independent binary variables that equal $+1$ with probability $F_{in}$ and $-1$ with probability $1-F_{in}$,
where $F_{in}$ denotes the input pond fraction.
Note that these variables are uncorrelated with
the $h_i$.
Following a random update sequence, Glauber dynamics eventually
yield a metastable state with
output pond fraction
$F_{out}$. Fig.~\ref{fig:MEPO_det_evo} shows the output
configurations with $F_{out}=0.15$, $0.30$, and $0.45$.
This metastability is consistent with previous findings
from a dynamical systems analysis \cite{SVG15_CNSNS}.

\begin{figure}
\centering
\begin{tabular}{cccc}
\includegraphics[width=0.24\textwidth]{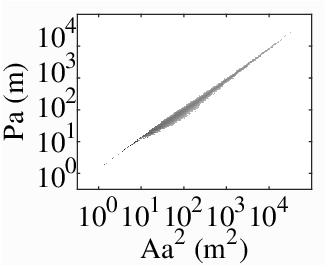} &
\includegraphics[width=0.24\textwidth]{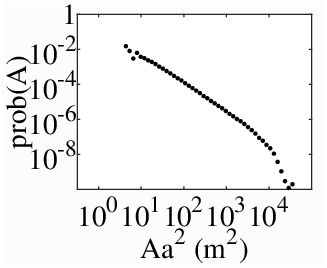} &
\includegraphics[width=0.21\textwidth]{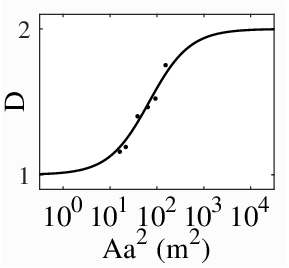} &
\includegraphics[width=0.21\textwidth]{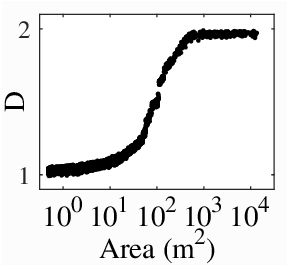} \\
(a) & (b) & (c) & (d)
\end{tabular}
\caption{\textbf{Geometrical characteristics of Ising model melt ponds.} Simulation data
in this figure are for the up-spin clusters in Fig.~\ref{fig:MEPO_det_evo}(c). (a) Log-log plot of the perimeter $P$ versus the area $A$, rendered as a (rescaled) density plot. (b) Log-log plot of the pond size distribution function $\textrm{prob}(A)$, with bin size 0.2 and very small ponds excluded. (c) Plot of the fractal dimension $D$ as a function of $A$ (log scale) for our melt pond Ising model. The individual points show the local fractal dimensions (computed from the lower edge of the convex hull of the data points in panel (a)) within the range $10$ m$^2$ $<$ $A$ $<$ $1,000$ m$^2$. (d) (Reproduced~\cite{HASPG12} with permission.) Plot of the fractal dimension as a function of area (log scale) based on image analysis of real melt ponds~\cite{HASPG12}. For panels (a)-(c), to compare with observations, $A$ and $P$ are shown in physical units with the lattice constant $a=0.85$ m, and the number of lattice sites is increased to $8192\times8192$ to improve the statistics.} \label{fig:MEPO_area_perim}
\end{figure}

The up-spin clusters in Fig.~\ref{fig:MEPO_det_evo}(c) at $F_{out}=0.45$ correspond to well developed melt ponds~\cite{HASPG12}. Fig.~\ref{fig:MEPO_area_perim}(a) shows the log-log plot of the perimeter $P$ versus the area $A$ for these clusters (shown in physical units as $Pa$ and $Aa^2$).
Fig.~\ref{fig:MEPO_area_perim}(b) shows the
pond size distribution function
$\textrm{prob}(A)$. It exhibits power law scaling $\textrm{prob}(A)\sim A^\zeta$ with the exponent
$\zeta=-1.58\pm 0.03$ for pond areas in the range $10$ m$^2$ $<$ $A$ $<$ $1,000$ m$^2$, in excellent agreement with the observed value
\cite{Perovich:JGR:2002} of about $-3/2$.

A key feature of multi-cluster systems is the tendency for smaller clusters to have simple shapes and larger clusters to have complex shapes. This onset of complexity can be quantified by an increase in the fractal dimension $D$, defined in terms of the perimeter $P$ and the area $A$ as $P\sim\sqrt{A}^D$. To find the critical area above which shapes do not remain simple, we choose the smallest possible $P$ for each $A$, or equivalently the lower edge of the cluster of points in the $(A,P)$-plane. Fig. \ref{fig:MEPO_area_perim}(c) shows the function $D(A)$ computed
for our model, illustrating
the fractal dimension transition from $1$ to $2$ around a critical area $A_c$.
Fitting a suitable smooth function to the data points \cite{Bowen_JFG_2016}, we find that the transition happens around the inflection point
$A_ca^2\approx70$ m$^2$. This predicted value agrees well with the
observed value~\cite{HASPG12} of about 100 m$^2$, as reproduced in Fig.~\ref{fig:MEPO_area_perim}(d).
The width of the transition regime in $\log{(A)}$
in Fig. \ref{fig:MEPO_area_perim}(c) also agrees well
with Fig. \ref{fig:MEPO_area_perim}(d).
Finally, Supplementary Fig.~\ref{fig:MEPO_elas} displays another quantifier
of the onset of complexity that accounts for the entire cluster of points in the $(A,P)$-plane. It yields the same critical area as before, $A_ca^2\approx70$ m$^2$.

Minimal models such as the RFIM necessarily have limitations. In particular, the RFIM has a percolation threshold very close to 0.5 at $H=0$ (see Supplementary Methods). This threshold decreases as $H$ decreases, but likely always exceeds the value for real melt ponds. This discrepancy may be attributed to unresolved processes at smaller scales, and/or the observed pre-melt ice topography being spatially correlated rather than completely random (see Supplementary Fig.~\ref{fig:psd}).
We anticipate that, based on a significant amount of observational data, a detailed scheme for choosing the initial spin configuration and update sequence may be formulated. See also Supplementary Methods for possible modifications to the update rule with an alternative
free energy yielding similar predictions.

The interpretation of complex Arctic melt ponds in terms of a simple disordered system may well advance our ability to model the future trajectory of the Arctic sea ice pack, e.g., through parameterizations in global climate models~\cite{FlFeTu10}. In addition, the statistical physics approach developed here may be generalizable to other systems near the transition point between ice and water, such as permafrost tundra lakes~\cite{SuVa14}.

\section*{Acknowledgments}
We gratefully acknowledge support from the Division of Mathematical
Sciences and the Division of Polar Programs at the U.S.
National Science Foundation (NSF) through Grants
DMS-1009704, ARC-0934721, DMS-0940249, DMS-1413454,
and DMS-0940262.
We are also grateful for
support from the Office of Naval Research (ONR) through Grant N00014-13-10291. Y.M. acknowledges support from a Vice Chancellor's Research Fellowship at Northumbria University. I.S. acknowledges support from the RFBR under the Grant \#16-31-60070 mol\_a\_dk. Finally, we would like to thank the NSF Math Climate Research Network (MCRN) for their support of this work.

\vspace{1ex}

\noindent
Y.M., I.S., C.S., and K.G.~proposed the model. Y.M. and C.S.~performed the numerical work. All authors contributed significantly to writing the manuscript.

\section*{Competing financial interests}
The authors declare no competing financial interests.

\newpage
\section*{Supplementary Methods}
\renewcommand{\figurename}{Supplementary Figure}
\renewcommand{\theequation}{S\arabic{equation}}
\setcounter{figure}{0}
\setcounter{equation}{0}

\begin{figure}[h]
\vspace{0.5cm}
\centerline{\includegraphics[width=0.5\textwidth]{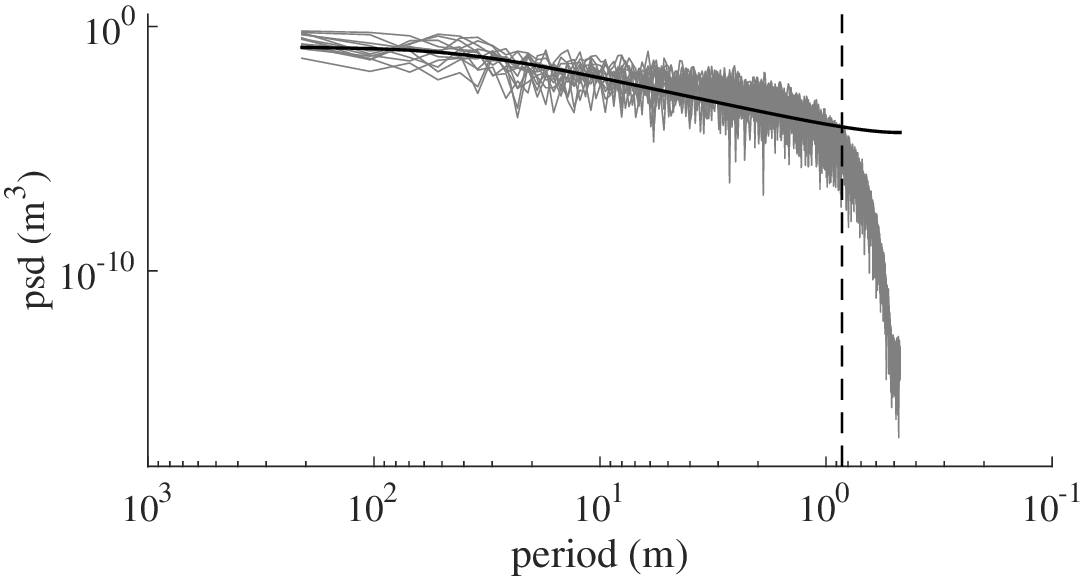}}
\caption{Snow depth power spectral density (gray curve) with corresponding null red noise spectrum (black curve). The lattice constant $a=0.85$ m is indicated by a vertical dashed line.}
\label{fig:psd}
\end{figure}

{\bf Lattice constant.} The lattice constant $a$ must be small relative to the 10-20 m length scales prominent in sea ice and snow topography\citeR{Petrich2012}. We set \mbox{$a=0.85$ m} as the length above which the power spectral density (psd) of observed snow topography exceeds a null red noise spectrum (Supplementary Fig.~\ref{fig:psd}). For this calculation, we used 13 radar transects collected during the Surface Heat Budget of the Arctic Ocean (SHEBA) project~\protect\citeR{Sturm2002}. To estimate the psd via the Welch modified periodogram, we calculated the power spectrum for each transect with a Hanning window and 50\% segment overlap, and then averaged the results across the transects. We calculated the corresponding null red noise spectrum based on lag-one spatial autocorrelation\citeR{Gilman1963} averaged across the transects.

\vspace{1ex}

\noindent
{\bf Alternative quantifier of the onset of complexity.} To account for the entire cluster of points in the $(A,P)$-plane in Fig.~\ref{fig:MEPO_area_perim}(a), we define a new quantifier of the onset of complexity as the variance $\sigma$ of $\log{(P)}$, hereafter referred to as the \textit{elasticity}. As shown in Supplementary Fig.~\ref{fig:MEPO_elas}, there exists a critical area $A_c$ such that $\sigma(\log{(P)})$ increases with $\log{(A)}$ for simple ponds with $A<A_c$, and decreases with $\log{(A)}$ for complex ponds with $A>A_c$. The onset of complexity may then be identified with maximum elasticity, which occurs at $A_ca^2\approx70$ m$^2$. This coincides with the critical area determined from Fig.~\ref{fig:MEPO_area_perim}(c) by the inflection point
in the best fit.

\begin{figure}[h]
\vspace{0.5cm}
\centerline{\includegraphics[width=0.44\textwidth]{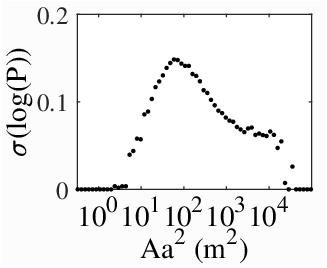}}
\caption{Plot of the variance $\sigma(\log{(P)})$ as a function of $A$ (log scale),
with bin size 0.2. The maximum happens at $A_ca^2\approx70$ m$^2$.}
\label{fig:MEPO_elas}
\end{figure}

\vspace{1ex}

\begin{figure}
\centering
\includegraphics[width=0.7\textwidth]{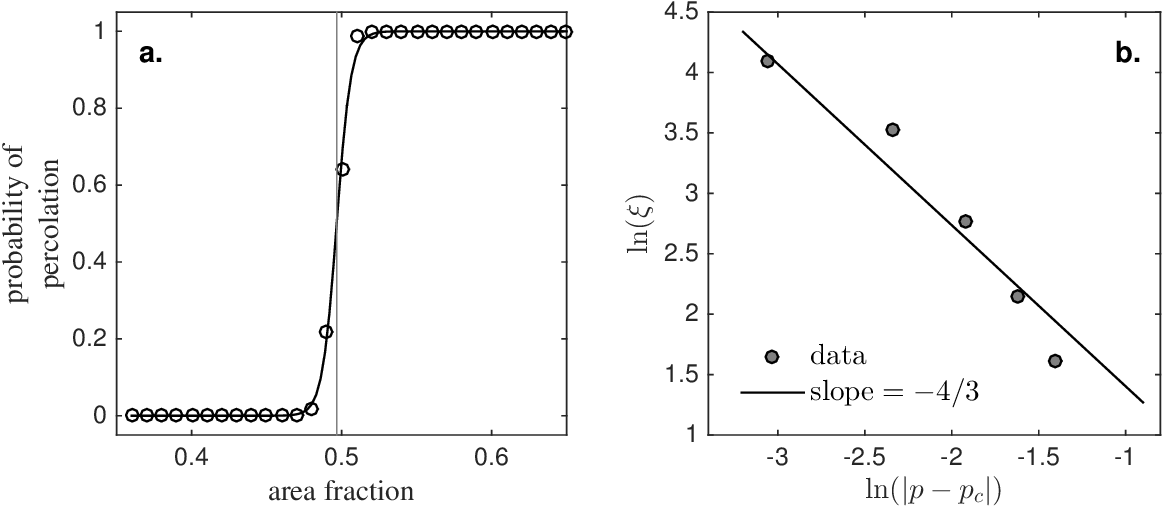}
\caption{(a) Probability of percolation as a function of area fraction. The curve is a hyperbolic tangent fit with inflection point close to 0.5 indicating the percolation threshold $p_c$. (b) Comparison of output from the Ising model (filled circles)
to the line with slope $-\nu=-4/3$ given by the universal
correlation length exponent $\nu$.}
\label{fig_crit}
\end{figure}

\noindent
{\bf Percolation threshold and correlation length exponent.} For a two dimensional square lattice with occupation probability $p$, the site-site correlation function $g(r_i,r_j)$ gives the probability that a site at $r_j$ is a member of the same cluster as a site at $r_i$. The function $g$ is assumed
to decay with large distance \mbox{$d=|r_i-r_j|$} according to
\begin{equation}
g(d)\sim \exp\left(-\frac{d}{\xi(p)}\right),
\label{eq_g}
\end{equation}
where $\xi(p)$ is referred to as the correlation length. Theory indicates that $\xi(p)$ should obey
\begin{equation}
\ln\xi(p)\sim -\nu\ln(|p-p_c|), \;\;\;\; p \longrightarrow p_c^-,
\end{equation}
where $\nu=4/3$ is the universal critical exponent in two dimensions and $p_c$ is the percolation threshold.
For the two-dimensional square site lattice,
$p_c \approx 0.59274621$ \citeR{Newman2000}. For the RFIM, analysis of 5,000 model realizations on $1024\times 1024$ lattices yields a value close to $p_c=0.5$ (Fig.\ \ref{fig_crit}a), with correlation lengths aligning reasonably with the universal exponent $\nu=4/3$ (Fig.\ \ref{fig_crit}b). This result indicates that the spatial correlation structure of melt ponds in this model is sufficiently short-ranged so that the system falls within a standard universality class \citeR{Isi92}.

\vspace{1ex}

\noindent
{\bf Time scale.} The time scale for melt pond formation can be generally
identified with the typical time taken to flip a spin in Glauber dynamics.
After the RFIM decides that a spin flip is energetically favorable,
we assume that the actual spin-flip process is facilitated by radiation
balance \citeR{Pie10_Book}. The incoming shortwave
radiation is $\mathrm{ISR}=Q(1-\alpha)$,
where $Q=$$\mathrm{460\; W\cdot m^{-2}}$ is the mean solar
insolation during polar summer, and $\alpha$ is
the surface albedo, $0.1$ for water and $0.5$ for ice.
The outgoing longwave radiation is $\mathrm{OLR}=\sigma T^4$,
where now $\sigma=$$\mathrm{5.67\times10^{-8}\; W\cdot m^{-2}K^{-4}}$
is the Stefan-Boltzmann constant, and $T$ is the surface temperature,
approximately $273$ K for both water and ice.
Therefore the rate of heat loss for ice
is $R_i=\mathrm{OLR}-\mathrm{ISR}=$$\mathrm{85\; W\cdot m^{-2}}$,
and the rate of heat gain for water
is $R_w=\mathrm{ISR}-\mathrm{OLR}=$$\mathrm{99\; W\cdot m^{-2}}$.
On the other hand, the energy per unit area required for freezing
a water column or melting an ice column is $E=L\rho h$,
where $L=$$\mathrm{3.34\times10^5\; J\cdot kg^{-1}}$ is the
latent heat of fusion, $\rho=$$\mathrm{1\times10^3 \; kg\cdot m^{-3}}$
is the density of water or ice (taken to be the same for simplicity),
and $h=$$\mathrm{0.3\; m}$ is a realistic value for the
height of the active layer.
Therefore, the time intervals needed to freeze a water site or to melt
an ice site are, respectively, $t_{w\rightarrow i}=E/R_i=14$ days
and $t_{i\rightarrow w}=E/R_w=12$ days, both of which
are reasonable, given this rough estimation.

\vspace{1ex}

\noindent
{\bf Nonzero uniformly applied field.} Let us choose $H\neq0$ and keep $J\rightarrow+\infty$ in the RFIM given by Eq.~(\ref{eq:H-RFIM}). Then the tiebreaker rule for a chosen site $i$ changes to $s_i=+1$ if $h_i<H$, and $s_i=-1$ if $h_i>H$, which favors ice for $H<0$ and water for $H>0$. Here we only consider two limiting cases when the tiebreaker rule completely favors ice or water: (I) $0\ll -H\ll J$; (II) $0\ll H\ll J$. In these cases, the random field $h_i$ does not affect the kinetics, so the RFIM reduces to the classical Ising model without disorder,
\begin{equation}\label{eq:H-CLIM}
\mathcal{H}=-H\sum_is_i-J\sum_{\langle i,j\rangle}s_is_j.
\end{equation}
The corresponding metastable states are known as Wulff droplets \citeR{SS98_CMP}. In case (I) the up-spin clusters are more elongated, and the percolation threshold is below 0.5. In case (II) the up-spin clusters are more circular, and the percolation threshold is above 0.5. These geometrical features afforded by varying $H$ (and possibly also $J$) provide additional prospects to describe detailed shapes of real melt pond patterns.

\vspace{1ex}

\noindent
{\bf Alternative update rule and free energy.} Let us retain the RFIM update rule when a majority exists among the neighboring sites, but adopt the following tiebreaker rule: the chosen site is updated to ice if its pre-melt ice height is larger than the average between the two neighboring ice sites, and water otherwise. For example, in Fig.~\ref{fig:photo}(b) we require that $s_P=+1$ if $h_P<(h_B+h_C)/2$, and $-1$ otherwise. This new update rule can be restated as minimizing an interfacial energy between water and ice: if a water site $i$ neighbors an ice site $j$, then a penalty $W-h_j$ is imposed, where $W\gg0$ is a constant. The total free energy $\mathcal{H}$ can then be written in two equivalent forms,
\begin{equation}\label{eq:H-MPIM}
\mathcal{H}=\sum_{\substack{\langle i,j\rangle:\\s_i>0,\,s_j<0}}\left(W-h_j\right)
\equiv\sum_is_i\Delta_ih-\sum_{\langle i,j\rangle}\frac{1}{2}s_is_j(W-\Omega_{ij}h),
\end{equation}
where $\Delta_i$ and $\Omega_{ij}$ represent, respectively, the discrete Laplacian at site $i$ and the average between sites $i,j$,
\begin{equation}
\Delta_ih\equiv h_i-\frac{1}{4} \sum_{j:\langle i,j\rangle} h_j,\quad
\Omega_{ij}h\equiv\frac{1}{2}(h_i+h_j).
\end{equation}
The new ``effective'' random fields $\Delta_ih$, being the curvature of $h_i$, are more correlated than the $h_i$ by themselves. As a result, at output pond fraction $F_{out}=0.45$, the critical area for the transition in fractal dimension and the critical area for maximum elasticity are both $A_ca^2\approx90$ m$^2$.
The corresponding power law scaling exponent for the
pond size distribution is $\zeta=-1.57\pm 0.03$. It may be interesting to compare these geometrical characteristics with classical ferromagnetic random bond Ising models \citeR{JaCo87_PRB}.

\bibliographystyleR{naturemag_noURL}

\end{document}